\documentclass[twocolumn,prl,showpacs,amsmath,amssymb]{revtex4}
\usepackage{graphicx}
\usepackage{amssymb,amsthm,amsfonts,amstext}
\usepackage{url}
\def\be{\begin{equation}}
\def\ee{\end{equation}}
\def\bea{\begin{eqnarray}}
\def\eea{\end{eqnarray}}
\def\bma{\begin{mathletters}}
\def\ema{\end{mathletters}}

\def\0{\overline{0}}

\def\q0{\underline{0}}

\def\H{{\cal H}}

\def\id{{\mathbb I}}

\def\H{{\cal H}}

\def\Z{\mathbb{Z}}

\def\tr{\mbox{tr}}
\def\one{\leavevmode\hbox{\small1\normalsize\kern-.33em1}}

\def\bra#1{\langle#1|} \def\ket#1{|#1\rangle}
\def\braket#1#2{\langle#1|#2\rangle}

\def\proj#1{\ket{#1}\!\bra{#1}}

\def\id{{\mathbb I}}
\def\tr{\mbox{tr}}

\begin{document}

\title{Quantum Steering and Space-Like Separation}

\author{Miguel Navascu\'es$^1$ and David P\'erez-Garc\'ia$^2$} 
\affiliation{$^1$School of Physics, University of Bristol, Bristol BS8 1TL, U.K.\\ $^2$Dpto. An\'alisis Matem\'atico and IMI, Universidad Complutense de Madrid, E-28040 Madrid, Spain}

\begin{abstract}
In non-relativistic quantum mechanics, measurements performed by separate observers are modeled via tensor products. In Algebraic Quantum Field Theory, though, local observables corresponding to space-like separated parties are just required to commute. The problem of determining whether these two definitions of `separation' lead to the same set of bipartite correlations is known in non-locality as \emph{Tsirelson's problem}. In this article, we prove that the analog of Tsirelson's problem in steering scenarios is false. That is, there exists a steering inequality that can be violated or not depending on how we define space-like separation at the operator level.
\end{abstract}

\maketitle

The concept of quantum steering was introduced by Schr\"{o}dinger \cite{schroedinger} in 1935 to analyze the EPR paradox \cite{EPR}. Roughly speaking, steering can be defined as the ability to transform the quantum state of a physical system by performing measurements in a distant laboratory. In the last years, we have witnessed a proliferation of results related to the quantum phenomenon of steering, both theoretical and experimental \cite{steering1,steering2,steering3,steering4,steering_exp1,steering_exp2,steering_exp3}. 

In all such works it is assumed that measurement operators corresponding to space-like separated observations act over different Hilbert spaces. A bipartite scenario is thus modeled via the tensor product of two Hilbert spaces, one for each (separate) degree of freedom. However, there is another way of modeling space-like separation; namely, by demanding separate measurement operators to commute. The problem of deciding if both definitions of space-like separation generate the same sets of bipartite correlations is known in non-locality as \emph{Tsirelson's problem}, and has recently generated considerable activity in Quantum Information Theory \cite{tsirel1,tsirel2,tsirel3, tsirel4,tsirel5}. Tsirelson's problem has been linked to the Connes embedding conjecture \cite{tsirel2,tsirel3}, and its negative resolution would have a tremendous impact in the von Neumann algebra community. In view of the importance of the problem and the formal relation between non-locality and steering, it is certainly remarkable that so far no one has ever considered how Tsirelson's non-locality problem translates to the steering arena, and what its resolution could be.

In this article, we show that, contrary to the non-locality case, the analog of Tsirelson's problem in steering scenarios \emph{can} be solved. Moreover, against all bets, the answer is negative! That is, there exist steering scenarios where the way we model space-like separation can lead to different experimental predictions. We prove this by defining a steering protocol where an untrusted party constrained by the tensor assumption cannot do better than using classical strategies. However, an untrusted party limited by commutation relations alone would be able to violate the associated steering inequality maximally. The fact that both models of space separation predict the same statistics in any finite dimensional setting makes this result extremely counterintuitive.

The structure of this article is as follows: first, we will discuss the concept of steering as it is usually presented in literature. Afterwards, we will introduce Tsirelson's problem and its analog in steering scenarios. Then we will present the main result of the paper: namely, we will derive a steering inequality that can only be violated if operator locality is defined via commutation relations. We will explain the connection between such a violation and the so-called heat vision effect \cite{heat_vision}, and, finally, we will present our conclusions.

\vspace{10pt}

In usual formulations of steering, two distant parties, Alice and Bob, share an unknown quantum state. Each of them is able to measure its respective subsystem in $s$ different ways, and such measurements are assumed to return one of $d$ possible outcomes. By choosing their measurements randomly and repeating the experiment many times, Alice and Bob are thus able to estimate each of the probabilities $P(a,b|x,y)$, i.e., the probability that Alice (Bob) observes the result $a$ ($b$) when she (he) performs the interaction $x$ ($y$). 

\begin{figure}
  \centering
  \includegraphics[width=8 cm]{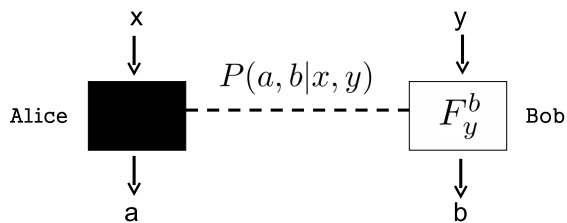}
  \caption{\textbf{Sketch of a steering protocol}. Two distant parties generate a set of bipartite correlations. While Alice's operations are not disclosed, Bob's measurement devices are assumed to be described by the operators $\{F^b_y\}\subset B(\H_B)$.}
  \label{proto}
\end{figure}

Up to this point, this symmetric scenario is identical to that used in non-locality. In steering, though, there is an extra premise: we assume that Bob's measurement devices are trusted, i.e., that Bob knows which measurement operators $\{F^b_y\}_b\subset B(\H_B)$ describe each measurement setting $y$. Alice's technological capabilities are, on the other hand, unknown: we have no information about her experimental setup (see fig. \ref{proto}). 

From all this it follows that the probabilities $P(a,b|x,y)$ will thus be given by

\be
P(a,b|x,y)=\tr(\sigma E^a_x\otimes F^b_y),
\ee

\noindent where both the quantum state $\sigma\in B(\H_A\otimes\H_B)$ and Alice's measurement operators $\{E^a_x\}\subset B(\H_A)$ are unknown.




Depending on the values of $P(a,b|x,y)$, Bob will decide if Alice is actually influencing his states by quantum means (i.e., by measuring her subsystem), in which case we speak of \emph{steering}, or, on the contrary, the statistics he observes are compatible with Alice having a classical knowledge of Bob's partial state. Indeed, it could be that, at each instance of the experiment, Bob's partial state is determined by some random variable $\lambda$ known to Alice, and each time she is asked to perform a measurement $x$, she gives a reply according to some probabilistic function of $\lambda$ and $x$. That would imply that $P(a,b|x,y)$ can be expressed as $P(a,b|x,y)=\int P(\lambda)d\lambda P(a|\lambda,x)\tr(F^b_y\rho_\lambda)$. If such is the case, we will say that the distribution $P(a,b|x,y)$ admits a \emph{local hidden state} (LHS) model \cite{steering1,steering2}.



In the previous (standard) formulation, we are implicitly assuming that the measurements conducted by Alice are associated with operators of the form $A\otimes \id_B$, while Bob's measurement operators can be expressed as $\id_A\otimes B$. This is indeed the way separate measurements are modeled in non-relativistic quantum mechanics. In Algebraic Quantum Field Theory (AQFT), though, measurements corresponding to causally unconnected regions of space-time are just required to commute \cite{AQFT}. That is, the conditions $\tilde{A}=A\otimes \id_B$, $\tilde{B}=\id_A\otimes B$ are relaxed to $[\tilde{A},\tilde{B}]=0$. Most authors in AQFT go further and also demand the \emph{statistical independence} of space-like separated regions \cite{yngvason}, i.e., the property that, for every pair of local states $\omega_A,\omega_B$, there exists a joint state $\omega_{AB}$ such that $\omega_{AB}(\tilde{A}\tilde{B})=\omega_A(\tilde{A})\omega_B(\tilde{B})$, for any pair of local operators $\tilde{A},\tilde{B}$. This can be shown equivalent, again, to Alice and Bob's operators being separated by tensor products. 

It is though far from clear that statistical independence is a fundamental physical property, and so it may well be that our universe violates it. As a matter of fact, in Haag and Kastler's seminal paper \cite{haag} on AQFT, a much weaker notion of statistical independence, namely \emph{essential uncoupling}, is only assumed to hold for infinitely distant regions of space-time. It therefore follows that commutation relations are as serious a candidate to model space-like separation as the tensor assumption. 

Tsirelson's non-locality problem consists, precisely, in deciding if both descriptions of separate measurements are equivalent at the level of correlations, i.e., if any distribution of the form $P(a,b|x,y)=\tr(E^a_xF^b_y\rho)$, with $[E^a_x,F^b_y]=0$, can be approximated by distributions of the form $\tilde{P}(a,b)=\tr(\tilde{E}^a_x\otimes \tilde{F}^b_y\tilde{\rho})$. It can be proven that both models of space separation lead to the same set of correlations if all measurement operators involved are assumed to act over finite-dimensional Hilbert spaces \cite{tsirel1,tsirel4}. Tsirelson's problem thus reduces to finding out if both sets of correlations also remain the same in infinite dimensions.

The generalization of Tsirelson's problem to steering scenarios is straightforward: once Bob's operators $\{F^b_y\}\subset B(\H_B)$ are given, we can model Alice's $\{E^a_x\}\subset B(\H_A\otimes \H_B)$ either by demanding them to admit an expression of the form $\tilde{E}^a_x\otimes \id^{(1)}_{B}$, where $\H^{(1)}_{B}$ denotes the part of $\H_B=\H^{(1)}_{B}\otimes \H^{(2)}_{B}$ where Bob's operators act non-trivially (the tensor assumption), or by imposing commutation relations of the type $[E^a_x,\id_A\otimes F^b_y]=0$ (the commutation assumption) \footnote{Note that, in both cases, representing the operations conducted by Alice \emph{and} Bob involves extending Bob's initial Hilbert space $\H_B$ to $\H_A\otimes \H_B$ and redefining Bob's measurement operators to act trivially on the extension $\H_A$.}. In this context, Tsirelson's steering problem would hence be to determine if there exists a steering protocol where the tensor and the commutation assumptions are experimentally distinguishable. In other words: can all bipartite distributions of the form $P(a,b|x,y)=\tr\{\sigma E^a_x F^b_y\}$, with $[E^a_x,F^b_y]=0$, be approximated by distributions of the form $\tilde{P}(a,b|x,y)=\tr\{\tilde{\sigma}\tilde{E}^a_x\otimes F^b_y\}$?



In \cite{tsirel4} it was shown that the answer to this question is positive if either $s=d=2$, or the initial set of measurements $\{E^a_x\}_{x=1}^s$ does not allow Alice to induce \emph{heat vision} in the joint system \cite{heat_vision}. The heat vision effect refers to the phenomenon that certain collections of von Neumann measurements can bring the system to a non-convergent dynamics when applied randomly and sequentially. In \cite{heat_vision}, it was shown that such a condition is equivalent to the ability of such measurements to induce an arbitrarily high energy increase in the system, for all reasonable definitions of energy (hence the name `heat vision').

Define the quantum channel $\omega_s$ that describes the process of Alice randomly measuring the system in one or other basis $x=1,...,s$. The idea of the proof in \cite{tsirel4} was to show that, if $\lim_{N\to\infty}\Omega^N(E^a_x\sigma E^a_x)=:\sigma^a_x$ exists for all $x,a$, then the state $\sum_a \sigma^a_x$ is independent of $x$. We would thus end up with a set of states $\{\sigma^a_x\}$ such that $P(a,b|x,y)=\tr(\tilde{\sigma}^a_xF^b_y)$, with $\sum_a\sigma^a_x=\hat{\sigma}$. As proven in \cite{HJW93,scho36, tsirel3, tsirel4}, this implies that there exists a state $\tilde{\sigma}$ and measurement operators $\tilde{E}^a_x$ with $P(a,b|x,y)=\tr\{\tilde{\sigma}\tilde{E}^a_x\otimes F^b_y\}$.

The fact that heat vision only arises in infinite dimensional systems means that, in scenarios where Bob's operators $\{F^b_y\}$ act over a finite-dimensional Hilbert space $\H_B$, both definitions of space-like separation lead to the same set of bipartite correlations. 

From all the above, it follows that a negative resolution of Tsirelson's steering theorem would thus involve defining a steering scenario with $\mbox{dim}(\H_B)=\infty$ and $sd>4$. We will do this soon, but first we have to introduce a couple of mathematical concepts.

Given two abstract groups $G_1,G_2$, the \emph{free product} $G_1*G_2$ is defined as a group whose generators are the generators or the groups $G_1,G_2$, and the product between generators $g_1\in G_1,g_2\in G_2$ belonging to different groups is regarded as a new element $g_1g_2$. The elements of $G_1*G_2$ are thus either the identity or finite `words' of alternate elements of $G_1$ and $G_2$.

For instance, take two groups isomorphic to $\Z_2$, with generators $g_1$ and $g_2$; the elements of each group are thus $\{1,g_1\}$ and $\{1,g_2\}$, respectively. Then the elements of the free product $\Z_2*\Z_2$ are generated by multiplying $g_1$'s and $g_2$'s according to the prescriptions $g_1\cdot g_2\equiv g_1g_2$, $g_2\cdot g_1\equiv g_2g_1$ and the property $g_x^2=1$. The first elements of $\Z_2*\Z_2$ are therefore $1$, $g_1$, $g_2$, $g_1g_2$, $g_2g_1$, $g_1g_2g_1$, ...

Given any group $G$, we can associate to it a Hilbert space $l_2(G)$ by defining an orthonormal basis labeled by the elements of $G$, i.e., $\{\ket{g}:g\in G\}$. An arbitrary vector $\ket{\Psi}\in l_2(G)$ can thus be expressed as $\ket{\Psi}=\sum_{g\in G}c_g\ket{g}$. The \emph{left regular representation} of $G$ is the set of unitary operators $\{\lambda_G(g):g\in G\}\subset B(l_2(G))$ that act over the basis elements of $l_2(G)$ according to the relation:

\be
\lambda_G(g)\ket{h}=\ket{gh},\forall h\in G.
\label{l_regular}
\ee

Consider then the group $G[s]=\overbrace{\Z_2*...*\Z_2}^{s\mbox{ times}}$, together with the operators $S_y\equiv \lambda_G(g_y)\in B(l_2(G[s]))$, for $y=1,...,s$. Note that, for $s\geq 2$, $G[s]$ has infinite order, and therefore $l_2(G[s])$ is infinite dimensional. 

From (\ref{l_regular}), it is immediate that $S_y^2=\id$ for all $y$, and so, for each $y$, the operators $(\id\pm S_y)/2$ are orthogonal projectors. 

Now, imagine a setup where both parties can perform $s$ different measurements of $d=2$ possible outcomes, with values $a,b\in\{-1,1\}$. Taking Bob's measurement operators to be $F_y^b\equiv (\id+bS_y)/2$, our steering scenario is completely defined. Consider then the following steering Bell-type functional:

\be
f_s\equiv \frac{1}{s}\sum_{y=1}^s P(a=b|y,y)-P(a\not=b|y,y)
\ee

\noindent Intuitively, $f_s$ measures how correlated Alice's and Bob's observations are when they choose the same inputs.

Let us estimate the maximum value $f^\star_s$ of such a functional under the assumption that Alice's and Bob's operators act over different Hilbert spaces, i.e., under the tensor assumption. Calling $\sigma$ the shared state and $\{E^a_x\}$ Alice's measurement operators (which can be assumed projectors), we have that

\be
f^\star_s=\max_{\sigma, \{R_x\}}\frac{1}{s}\sum_{y=1}^s\tr(\sigma R_y\otimes S_y) =\max_{\{R_x\}}\left\|\frac{1}{s}\sum_{y=1}^s R_y\otimes S_y\right\|,
\ee

\noindent where each operator $R_x\equiv E^{a=1}_x-E^{a=-1}_x$, like the $\{S_y\}$'s, represents an observable with spectrum in $\{-1,1\}$ \footnote{It is easy to see that $f^\star_s$ also bounds the value of $f_s$ in more complicated situations where Bob's operators are of the form $B=\oplus_{i=1}^\infty \id^{(1)}_{B(i)}\otimes \tilde{B}_i\in B(\H_B)$ and Alice's operators are just allowed to act non-trivially over the space $\H_A\otimes (\oplus_{i=1}^\infty \H^{(1)}_{B(i)})$}. 

For any word of the type $g=g_1g_3g_2g_1...$, denote by $R_g$ the operator $R_1R_3R_2R_1$..., and define the unitary operator $U\equiv\sum_{g\in G}R^{-1}_g\otimes\proj{g}$. Notice that $U\left(\sum_{y=1}^s R_y\otimes S_y\right)U^\dagger=\sum_{y=1}^s\id_A\otimes S_y$. It follows that

\be
f^\star_s=\left\|\frac{1}{s}\sum_{y=1}^s S_y\right\|.
\label{norm}
\ee

\noindent The above expression tells us that the given steering functional can be maximized if Bob's state is close to the maximal (minimal) eigenstate of the operator between the norm signs and Alice always outputs the result $+1$ ($-1$). Hence, under the tensor assumption, Bob will always find a LHS model compatible with the observed value of $f_s$.

In the Appendix, we show that the norm on the right hand side of eq. (\ref{norm}) is equal to $2\frac{\sqrt{s-1}}{s}$. That is,

\be
\frac{1}{s}\sum_{y=1}^s P(a=b|y,y)-P(a\not=b|y,y)\leq 2\frac{\sqrt{s-1}}{s},
\label{ineq}
\ee

\noindent under the tensor assumption. Note that $f^\star_s$ is strictly smaller than 1 for $s>2$. This means that, as long as there are three or more measurements in our steering scenario, Alice and Bob cannot be perfectly correlated. Also, note that $f^\star_s\to 0$ when $s\to\infty$, i.e., as $s$ grows, Alice and Bob become more and more uncorrelated.

To conclude our argument, we have to show that there exists a choice of dichotomic operators $\{R_y\}$, with $[R_y,S_x]=0$ that allows to violate eq. (\ref{ineq}). Now, in an analogous way in which we defined the left regular representation, given a group $G$, one can define its associated \emph{right regular representation} \cite{pisier} as

\be
\rho_G(g)\ket{h}=\ket{hg^{-1}}, \forall h\in G.
\ee

\noindent Note that the right regular representation commutes with the left regular. Indeed, $\rho_G(g)\lambda_G(g')\ket{h}=\ket{g'hg^{-1}}=\lambda_G(g')\rho_G(g)\ket{h}$. It follows that the dichotomic operators $\hat{R}_x\equiv \rho_{G[s]}(g_x)\in B(l_2(G[s]))$, for $x=1,...,s$, commute with $\{S_y\}_{y=1}^s$. Assume, thus, that Alice is measuring $\{\hat{R}_x\}_{x=1}^s$ and that Alice and Bob share the state $\sigma\equiv\proj{1}$. Then it is immediate that

\be
f_s=\frac{1}{s}\sum_{y=1}^s\tr(\sigma\hat{R}_yS_y)=1,
\ee

\noindent i.e., Alice and Bob can be perfectly correlated for all values of $s$, hence violating eq. (\ref{ineq}) for $s>2$. 

\noindent The assumptions of tensor structure and commutativity of separate measurements therefore lead to different results in steering scenarios. Tsirelson's steering problem has been solved.

The fact that the bound $f^\star_s$ was beaten means that the corresponding probabilities $P(a,b|x,y)$ do not admit a tensor representation with Bob's operators remaining the same. From \cite{tsirel4}, it thus follows that Alice's measurements must be able to induce heat vision. 

This is precisely what happens in this example. Indeed, define the quantum channel $\omega_s$ that describes the process of randomly measuring the system in one or other basis $x=1,...,s$. Seen as a superoperator $\bar{\omega}_s$ acting over the Hilbert space $l_2(G[s])\otimes l_2(G[s])$, $\omega_s$ has the form:

\be
\bar{\omega}_s=\frac{1}{2}\id\otimes\id+\frac{1}{2s}\sum_{x=1}^s\hat{R}_x\otimes \hat{R}_x^*.
\ee

\noindent By symmetry with the left regular representation, it is easy to see that $\|\bar{\omega}_s\|=\frac{1}{2}+\frac{f^\star_s}{2}$. Consequently, for any initial state $\rho\in B(l_2(G[s]))$, the purity of $\omega_s^N(\rho)$ will be upper bounded by $\left(\frac{1}{2}+\frac{f^\star_s}{2}\right)^{2N}$, i.e., it will tend to zero exponentially fast. This implies that $(\omega_s^N(\rho))$ tends to zero in the Hilbert-Schmidt norm, and thus has no limit in trace norm. In sum, Alice's measurements $\{\hat{R}_x\}_{x=1}^s$ induce heat vision.

\vspace{10pt}
\noindent\emph{Conclusion}

In this work, we have introduced and solved the analog of Tsirelson's problem in steering scenarios. Our surprising conclusion is that, contrary to the finite-dimensional case, they way local observables are modeled in steering protocols \emph{does} make a difference in infinite dimensions. This result has important consequences for the security analysis of semi-device independent protocols, like \cite{marcin}: indeed, since it is not known which of the two models of space-like separation holds in our world, in order to be on the safe side, it must be assumed that untrusted parties are limited by commutation relations alone. It is therefore a fortunate coincidence that multipartite correlations defined via commutation relations are precisely the ones that we know how to characterize nowadays \cite{hier1,hier2}.



\noindent\emph{Acknowledgements}

M. N. has been supported by the Templeton Foundation. D.P.-G. acknowledges QUEVADIS, CQC and Spanish grants QUITEMAD, MTM2011-26912 and  PRI-PIMCHI-2011-1071. We thank Tobias Fritz for interesting discussions.

\section{Computation of $f_s^\star$}
\begin{appendix}
The aim of this Appendix is to prove that $f^\star_s$, as defined in eq. (\ref{norm}), is equal to $2\frac{\sqrt{s-1}}{s}$. This amounts to calculating the norm of the operator

\be
\Omega_s\equiv \frac{1}{s}\sum_{y=1}^sS_y.
\ee

Let $\ket{\Phi}$ be a normalized vector in $l_2(G)$ such that $|\bra{\Phi}\Omega_s\ket{\Phi}|>\|\Omega_s\|-\epsilon$. It is easy to see that we can choose $\ket{\Phi}$ such that $\braket{1}{\Phi}=0$: just realize that

\be
\bra{\Phi}\Omega_s\ket{\Phi}=\bra{\Phi}\rho_G(g)^\dagger\Omega_s \rho_G(g)\ket{\Phi},
\ee

\noindent for any $g\in G[s]$. If we can approximate $\ket{\Phi}$ in 2-norm up to precision $\delta$ with linear combinations of words $h$ of length $|h|=N$, then choosing $|g|>N$ guarantees that the vector $\rho_G(g)\ket{\Phi}$ will have an overlap with $\ket{1}$ of at most $\delta$.

We can thus assume $\ket{\Phi}$ to be of the form $\ket{\Phi}=\sum_{y=1}^s\ket{\Phi_y}$, with each vector $\ket{\Phi_y}$ being a linear combination of words starting with the letter $g_y$. Then one can check that

\be
s|\bra{\Phi}\Omega_s\ket{\Phi}|=\left|\sum_{y\not=z}(\bra{\Phi_y}S_y\ket{\Phi_z}+\bra{\Phi_z}S_y\ket{\Phi_y})\right|.
\label{cota_sup}
\ee

Define $p_y\equiv \|\ket{\Phi^y}\|^2$, and note that, for fixed $y$, the vectors $\{S_y\ket{\Phi_z}:z\not=y\}$ are orthogonal to each other. This implies that the vector $\ket{\Psi_y}\equiv\sum_zS_y\ket{\Phi_z}$ has norm equal to $\sqrt{1-p_y}$. Eq. (\ref{cota_sup}) can then be rewritten as

\be
\left|\sum_y(\braket{\Phi_y}{\Psi_y}+\braket{\Psi_y}{\Phi_y})\right|\leq 2\sum_{y=1}^s \sqrt{p_y}\sqrt{1-p_y}.
\ee

\noindent By the Cauchy-Swartz inequality, the last term can be upper bounded by $2\sqrt{s-1}$. 

Putting all together, we have that

\be
f^\star_s=\|\Omega_s\|\leq 2\frac{\sqrt{s-1}}{s}.
\ee

It just suffices to prove that the former upper bound is tight, i.e., that there exists a sequence of normalized vectors $(\ket{\varphi_N})$ such that 

\be
\lim_{N\to\infty}|\bra{\varphi_N}\Omega_s\ket{\varphi_N}|=2\frac{\sqrt{s-1}}{s}.
\ee

\noindent It can be checked that the family of vectors

\be
\ket{\varphi_N}\equiv \sum_{|g|\leq N}\frac{1}{\sqrt{N+1}\sqrt{f(|g|)}}\ket{g},
\ee

\noindent with $f(k)$ being the number of words of length $k$, does the trick.
\end{appendix}

\end{document}